\documentstyle[twoside,fleqn,npb,epsfig]{article}
\newcommand{\beq}{\begin{equation}}
\newcommand{\eeq}{\end{equation}}
\newcommand{\ba}{\begin{eqnarray}}
\newcommand{\ea}{\end{eqnarray}}

\newcommand{\bpt}{\bm p_T^{}}
\newcommand{\bkt}{\bm k_T^{}}

\newcommand{\bm}[1]{\mbox{\boldmath${#1}$}}

\newcommand{\AmS}{{\protect\the\textfont2
  A\kern-.1667em\lower.5ex\hbox{M}\kern-.125emS}}

\begin{document}

\begin{titlepage}

\centerline{\Large \bf Intrinsic transverse momentum and transverse spin
asymmetries}

\vspace{7mm}

\centerline{\bf Dani\"el Boer}

\vspace{4mm}

\centerline{RIKEN-BNL Research Center}
\centerline{Brookhaven National Laboratory, Upton, NY 11973, U.S.A.}

\vspace{30mm}
\centerline{\bf Abstract}
\vspace{4mm}
We investigate leading twist transverse momentum dependent origins of 
transverse spin asymmetries in hadron-hadron collisions. The chiral-odd T-odd 
distribution function with intrinsic transverse momentum dependence, which 
would signal an intrinsic handedness of quarks inside a hadron, could account 
for single spin asymmetries and at the same time for the large 
$\cos 2\phi$ asymmetry in the unpolarized Drell-Yan cross section, 
which still lacks understanding. We show explicitly how it would relate 
unpolarized and polarized observables measurable with proton-proton 
collisions at RHIC. It would offer 
a new possibility to access the transversity distribution function. 
\vspace{30mm}

\centerline{Talk given at the 7th International Workshop on} 
\centerline{``Deep Inelastic Scattering and QCD'' (DIS99)}
\centerline{DESY-Zeuthen, April 19 - 23, 1999}

\end{titlepage}

% declarations for front matter
\title{Intrinsic transverse momentum and transverse spin asymmetries}

\author{Dani\"el Boer\address{RIKEN-BNL Research Center,
Brookhaven National Laboratory, Upton, NY 11973, U.S.A.}}

\begin{abstract}
We investigate leading twist transverse momentum dependent origins of 
transverse spin asymmetries in hadron-hadron collisions. The chiral-odd T-odd 
distribution function with intrinsic transverse momentum dependence, which 
would signal an intrinsic handedness of quarks inside a hadron, could account 
for single spin asymmetries and at the same time for the large 
$\cos 2\phi$ asymmetry in the unpolarized Drell-Yan cross section, 
which still lacks understanding. We show explicitly how it would relate 
unpolarized and polarized observables measurable with proton-proton 
collisions at RHIC. It would offer 
a new possibility to access the transversity distribution function. 
\end{abstract}

\maketitle

\section{Introduction}
Large single transverse spin asymmetries have been observed 
in the process $p \, p^{\uparrow} \rightarrow \pi \, X$ \cite{Adams}.
Of course, one experiment only cannot reveal the origin(s) of such asymmetries
conclusively and one needs comparison to other experiments, for instance 
the planned RHIC spin physics experiments.

The transverse 
momentum dependence of transverse spin asymmetries should be related to the 
transverse momentum of quarks inside a hadron. Our goal is to 
investigate the relation between  
the transverse spin and transverse momentum of quarks. 

\begin{figure}[htb]
\vspace{-12pt}
\begin{center}
\includegraphics[width = 5.5 cm]{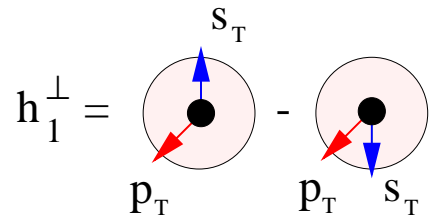}
\end{center}
\vspace{-22pt}
\caption{The chiral-odd T-odd 
distribution function $h_1^\perp$ ($s_T$ and $p_T$
are the quark's transverse spin and momentum).}
\label{h1p}
\end{figure}

In Ref.\ \cite{DB-99} 
we have argued that conventional perturbative QCD and higher
twist effects do not produce large --if any-- single transverse spin
asymmetries. Less conventional higher twist mechanisms, such as
soft gluon poles in twist-3 matrix elements \cite{QS-91b} or the effectively
equivalent twist-3 T-odd distribution functions \cite{Boer4}, could produce
a single spin asymmetry. 
For the Drell-Yan (DY) process it is expected to be similar 
in size to the double spin asymmetry $A_{LT}$ \cite{Jaffe-Ji-91}, 
which is estimated \cite{Kanazawa} to be much smaller than the double
transverse spin asymmetry $A_{TT}$, that is estimated to be on
the order of percents for RHIC energies \cite{Martin}. 
Therefore, we proposed
an alternative explanation of such a single spin asymmetry, involving a 
particular leading twist, 
intrinsic transverse momentum dependent, chiral-odd, T-odd distribution 
function \cite{Boer-Mulders-97}, called 
$h_1^\perp$ (cf.\ Fig.\ \ref{h1p}; depicted are probabilities of specific
quark states (black dot) inside a hadron). 
This can not only offer an explanation for single
spin asymmetries in $p \, p^{\uparrow} \rightarrow \pi \, X$ or 
the DY process, but also for the large azimuthal 
$\cos 2\phi$ dependence of the unpolarized DY cross section 
\cite{NA10,Conway}, which still lacks understanding.    

\begin{figure}[htb]
\vspace{-6pt}
\begin{center}
\includegraphics[width = 5.5 cm]{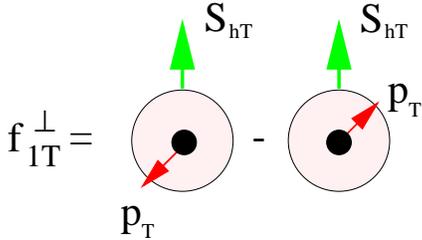}
\end{center}
\vspace{-22pt}
\caption{The chiral-even T-odd 
distribution function $f_{1T}^\perp$ ($S_{hT}$ is the hadron transverse spin).}
\label{f1Tp}
\end{figure}
Unlike its chiral-even counterpart $f_{1T}^\perp$ (investigated in
\cite{s90,Anselmino}), which depends on the polarization of the parent hadron
(cf.\ Fig.\ \ref{f1Tp}),
the function $h_1^\perp$ signals an {\em intrinsic handedness\/} inside an
{\em unpolarized\/} hadron. It would mean an 
orientation dependent correlation between the transverse spin and the
transverse momentum of quarks inside an unpolarized hadron\footnote{In Ref.\
\cite{DB-99} we have discussed 
the theoretical difficulties associated with T-odd
distribution functions.}.
One can
use the polarization of another hadron to become sensitive to the polarization
of quarks inside an unpolarized hadron. In this way it could provide a new 
way of 
measuring the transversity distribution function $h_1$. 
For this purpose we 
propose two measurements that could be done at RHIC 
using polarized proton-proton collisions.

\section{An unpolarized asymmetry}

\begin{figure}[htb]
\includegraphics[width = 7.2 cm]{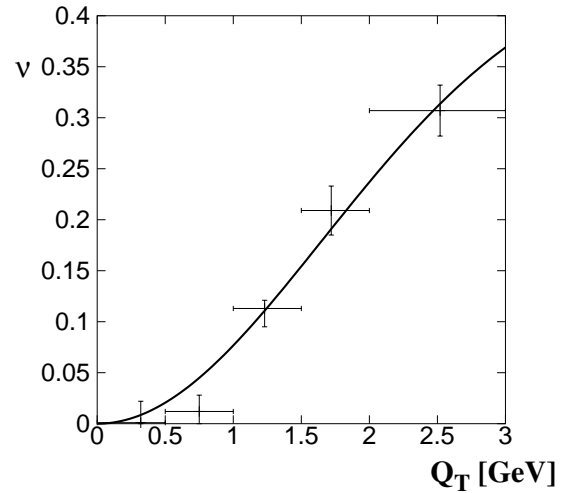}
\vspace{-12pt}
\caption{\label{fit}Data from \protect\cite{NA10} at 194 GeV and fit 
to $\nu$ as a function of the 
transverse momentum $Q_T$ of the lepton pair.}
\label{fit1}
\end{figure}

A large $\cos 2\phi$ angular dependence in the unpolarized DY process 
$\pi^- N \rightarrow \mu^+ \mu^- X$, where $N$ is either 
deuterium or tungsten and for instance using a $\pi^-$ beam of 194 GeV, 
was found by the NA10 Collaboration \cite{NA10}. The perturbative 
QCD prediction (NLO) for the cross section written as
\beq
\frac{d\sigma}{d\Omega} \propto 
1+ \cos^2\theta + \sin^2\theta \left[\mu  
\cos\phi + \frac{\nu}{2} {\cos 2\phi} \right],
\eeq
is $\mu \! \approx \! 0, \, \nu \!\approx \! 0$. However,
$\nu$ acquires values of more than 0.3 depending on the transverse momentum
$Q_T$ 
of the muon pair (its invariant mass is between 4 and 8 GeV/$c^2$), cf.\ Fig.\
\ref{fit1}. 
Even though the cross section itself is dependent on the nuclear target, since
{$\sigma_W(Q_T)/\sigma_D(Q_T)$} is an increasing function of 
$Q_T$, the analyzing power $\nu(Q_T)$ shows no apparent nuclear 
dependence, indicating that the asymmetry arises at the quark-hadron level. 

In Ref.\ \cite{DB-99} 
we have observed that within the framework of transverse momentum dependent 
distribution functions \cite{Tangerman-Mulders-95a,Boer-Mulders-97}, 
this asymmetry can only 
be accounted for by the
function $h_1^\perp$, unless $1/Q^2$ suppressed \cite{Berger-80}. 
Moreover, higher twist effects are expected to produce $\mu > \nu$, which is
not the case. We found   
$\nu \propto h_1^{\perp\, \pi} \, h_1^{\perp\, N}$ and used this
observation to fit the function $h_1^{\perp}$, assuming some simplifications, 
like independence of the type of parent hadron, cf.\ Fig.\ \ref{fit1}.  
In a similar way one can try to 
measure $\left<\cos 2\phi \right>$ in unpolarized 
$p \, p \rightarrow \mu^+ \mu^- X$ at RHIC and obtain a parametrization of 
$h_1^{\perp\, p}$. 

\section{Single spin asymmetries}

\begin{figure}[htb]
\includegraphics[width = 7.2 cm]{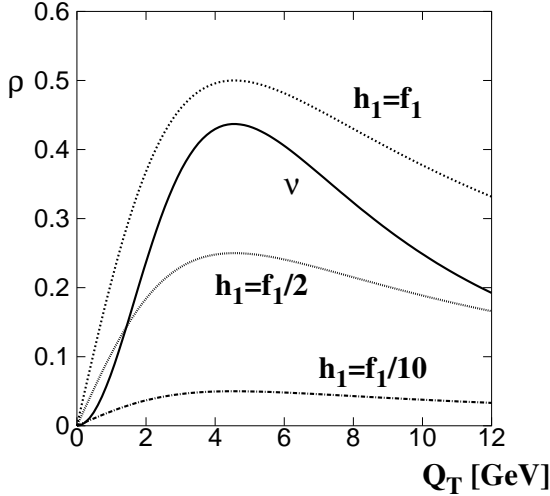}
\vspace{-12pt}
\caption{Upper bound predictions for $\rho$ 
for $h_1/f_1=\{1,1/2,1/10\}$, using the fit for $\nu$.}
\label{fit2}
\end{figure}

At RHIC they will also be able to measure 
$p \, p^\uparrow \rightarrow \mu^+ \mu^- X$. 
With one transversely polarized hadron the DY cross section 
will have more complicated azimuthal dependences. For instance 
(displaying only two terms):
\beq
\frac{d\sigma}{d\Omega \, d\phi_{S_T}} \propto 
\sin^2 \theta \left[ \frac{{\nu}}{2} \cos 2\phi - {\rho}\,
{\sin(\phi+\phi_{S_T})} \right].
\eeq
The analyzing power $\rho$ is proportional to the product 
$h_1^\perp \, h_1 $ \cite{DB-99}. 
Hence, the measurement of $\left<\cos 2\phi \right>$
combined with a measurement of the 
single spin azimuthal asymmetry $\left< {\sin(\phi+ \phi_{S_T})} \right>$ 
could provide information on $h_1$. In other words, 
a nonzero function $h_1^\perp$ will imply a relation between $\nu$ and $\rho$,
which in case of one flavor is ($\nu_{\mbox{${\rm max}$}}$ is the maximum
value attained by $\nu (Q_T)$)
\beq
\rho = 
\frac{1}{2} \;|\bm S_{T}^{}|\, 
\sqrt{\frac{{\nu} }{{\nu_{\mbox{${\rm max}$}}}}}\, \frac{h_1}{f_1}.
\eeq
This depends on the magnitude of $h_1$ compared to $f_1$ and in Fig.\
\ref{fit2} we display three options for $\rho$, using the fitted function 
$\nu$ which we view
as an optimistic upper bound. 

We note that the function $f_{1T}^\perp$ generates a totally 
different angular single spin asymmetry, namely 
$( 1+ \cos^2\theta )\;|\bm S_{T}^{}|\;
{\sin(\phi-\phi_{S_T})}\; {f_{1T}^{\perp}} \; f_1$.

The large single spin asymmetries found in $p \, p^{\uparrow} \rightarrow 
\pi X$ can also arise from leading twist T-odd functions with 
transverse momentum dependence. There are three options:
\ba
&& {h_1^\perp(x_1,\bpt)} \otimes h_1(x_2) \otimes D_1(z),\\
&& {f_{1T}^\perp(x_1,\bpt)} \otimes f_1(x_2)  \otimes D_1 (z),\\
&& h_1(x_1)  \otimes f_1(x_2)  \otimes H_1^\perp(z,\bkt).
\ea
The first two options are similar to the ones described above, accompanied
by the unpolarized fragmentation function $D_1$. The third option contains 
the Collins effect function $H_1^\perp$ \cite{Collins-93b}, which is formally 
the fragmentation function analogue of $h_1^\perp$, but in principle 
unrelated in magnitude.  
The last two options were investigated in \cite{Anselmino}.
 
\section{Conclusion}

The chiral-odd T-odd distribution function $h_1^\perp$ can not only offer an
explanation for single transverse spin asymmetries in
hadron-hadron collisions, but also for the unpolarized $\cos 2 \phi$ asymmetry 
in the $\pi^- N \rightarrow \mu^+
\mu^- X$ data (unlike any other function in this approach, unless $1/Q^2$ 
suppressed). It would 
relate unpolarized and polarized observables and thus would offer {a new
possibility to access $h_1$} in $p \, p \rightarrow \mu^+ \mu^- X$.

\end{document}